\documentclass[11pt]{article}

\usepackage{tikz}
\usetikzlibrary{shapes,arrows}
\usetikzlibrary{backgrounds, positioning, fit}
\usepackage{dblfloatfix}
\usepackage{booktabs}
\usepackage[inline]{enumitem}
\newcommand{\myNoSpacing}{
  \setlength{\itemsep}{0pt}
  \setlength{\parskip}{0pt}
  \setlength{\parsep}{0pt}
}
\usepackage{graphics}
\usepackage[T1]{fontenc}
\usepackage{xspace}
\usepackage{soul}
\usepackage{multirow}
\usepackage[table]{colortbl}
\usepackage{versions}
\usepackage{pifont}
\usepackage[hang,flushmargin]{footmisc}
\usepackage[normalem]{ulem}
\usepackage{enumitem}   
\usepackage[most]{tcolorbox}
\usepackage[colorlinks=true,citecolor=blue]{hyperref}
\usepackage{url}
\usepackage{amssymb,amsmath,amsthm}
\usepackage{thmtools}
\usepackage{algorithm}
\usepackage{algorithmic}
\usepackage[table]{xcolor}

\definecolor{blond}{rgb}{0.98, 0.94, 0.75}

\declaretheoremstyle[
    numbered = no,
    headfont = \bfseries,
    postheadspace = 0pt,
    headpunct = {}
]{thmsty}

\tcolorboxenvironment{takeaway}{
  enhanced jigsaw,
  colback     = blond,
  drop shadow,
  boxrule     = 0.9pt,
  boxsep      = 0.1pt,
  left        = 2pt,
  right       = 4pt,
  top         = 4pt,
  bottom      = 4pt,
}

\newcommand{\titlename}{\method: Mitigating PII Leakage in Language Models with \underline{P}rivacy-\underline{A}ware \underline{T}argeted \underline{C}ircuit Patc\underline{H}ing}

\newcommand{\method}{\textsc{Patch}\xspace}

\newcommand{\GS}{\textit{GPT2-Small}~}
\newcommand{\GM}{\textit{GPT2-Medium}~}
\newcommand{\GL}{\textit{GPT2-Large}~}
\newcommand{\LM}{\textit{Llama-3.2-1B}~}
\newcommand{\QB}{\textit{Qwen3-1.7B}~}
\newcommand{\QS}{\textit{Qwen3-0.6B}~}

\usepackage[final]{acl/acl}

\usepackage{times}
\usepackage{latexsym}

\usepackage[T1]{fontenc}

\usepackage[utf8]{inputenc}

\usepackage{microtype}

\usepackage{inconsolata}

\usepackage{graphicx}

\title{\titlename}

\author{Anthony Hughes$^1$, Vasisht Duddu$^2$, N. Asokan$^2$, Nikolaos Aletras$^1$, Ning Ma$^1$\\
$^1$University of Sheffield, $^2$University of Waterloo\\
\texttt{\{ajhughes3, n.ma, n.aletras\}@sheffield.ac.uk},\\ \texttt{vasisht.duddu@uwaterloo.ca}, \texttt{asokan@acm.org}\\
}

\begin{document}
\maketitle
\begin{abstract}
Language models (LMs) may memorize personally identifiable information (PII) from training data, enabling adversaries to extract it during inference. Existing defense mechanisms such as differential privacy (DP) reduce this leakage, but incur large drops in utility. 
Based on a comprehensive study using circuit discovery to identify the computational circuits responsible PII leakage in LMs, we hypothesize that specific PII leakage circuits in LMs should be responsible for this behavior. Therefore, 
we propose \method (\underline{P}rivacy-\underline{A}ware \underline{T}argeted \underline{C}ircuit Patc\underline{H}ing), a novel approach that first identifies and subsequently directly edits PII circuits to reduce leakage. \method achieves better privacy-utility trade-off than existing defenses, e.g., reducing recall of PII leakage from LMs by up to $65\%$. Finally, \method can be combined with DP to reduce recall of residual leakage of an LM to as low as $0.01\%$. Our analysis shows that PII leakage circuits persist even after the application of existing defense mechanisms. In contrast, \method can effectively mitigate their impact.\footnote{Code and data are publicly available at \url{https://github.com/ssg-research/pii-patch/}}
\end{abstract}

\myNoSpacing
\section{Introduction}
\label{sec:introduction}

Language models (LMs) have demonstrated remarkable advances \citep{gemmateam2024gemma2improvingopen, grattafiori2024llama3herdmodels}, yet their tendency to memorize training data poses privacy risks~\citep{kandpal2022deduplicating, NEURIPS2023_a1d20cc7,duan_uncovering_2024,hayes_measuring_2025}. In particular, prior work has shown that LMs can memorize and reproduce personally identifiable information (PII) from their training data~\citep{huang-etal-2022-large, kim2023propile,nakka-etal,borkar2025privacy}. This makes it possible for adversaries with black-box access to a model to expose such information \citep{lukas2023analyzing}.

\begin{figure}[!t]
    \centering
    \resizebox{.45\textwidth}{!}{
    \includegraphics[trim={1.5cm 0 0 0},scale=0.35]{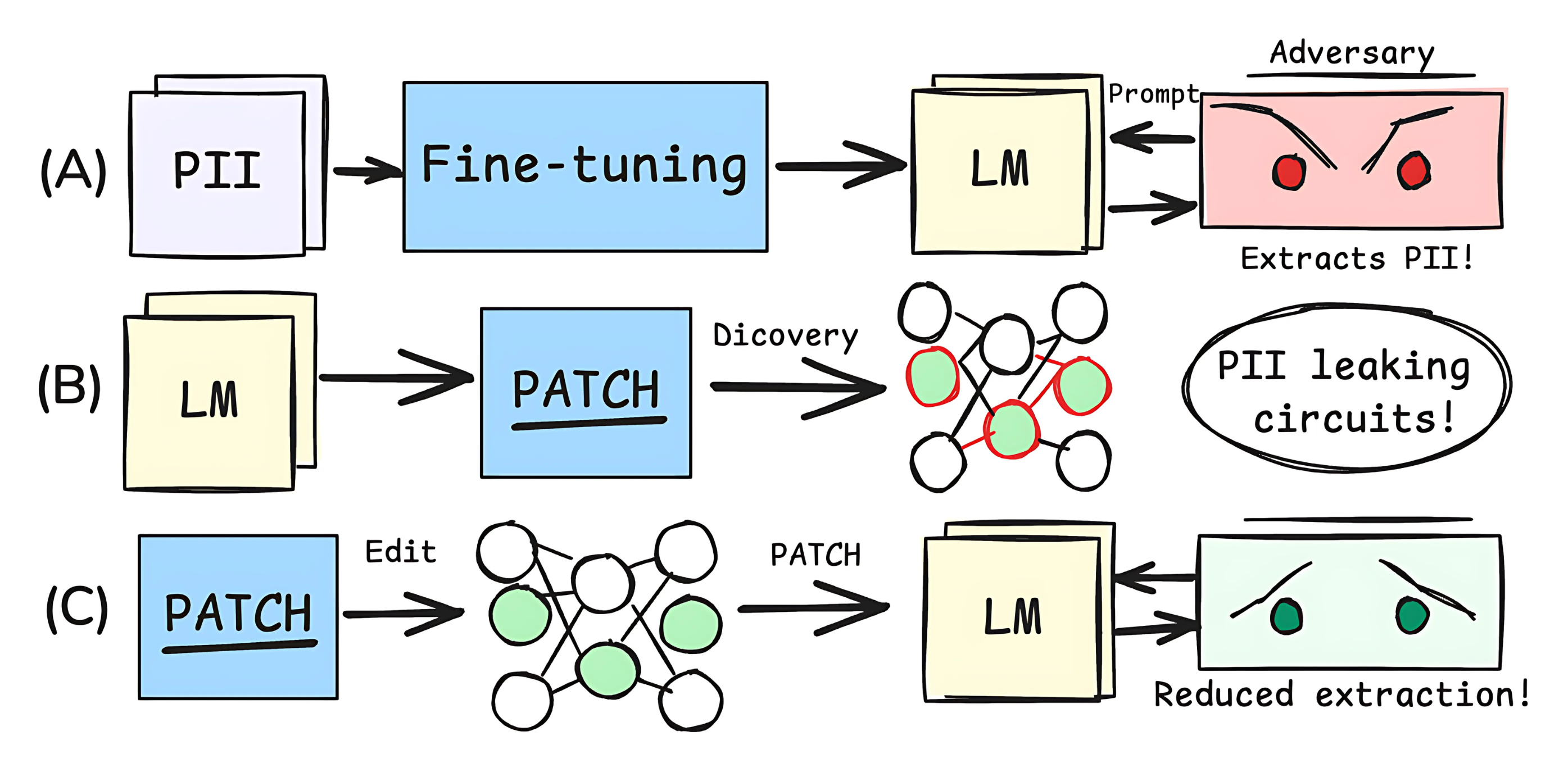}
    }
    \caption{Mitigating PII leakage through circuit analysis of LMs fine-tuned on PII-containing documents \textbf{(A)} with and without privacy defenses. \textbf{(B)} We use \method to discover PII leaking circuits. Finally, 
    \textbf{(C)} \method then edits the discovered circuits, reducing PII leaks.
    }
    \label{fig:paper-overview}
\end{figure}

A range of defense mechanisms have been proposed to mitigate PII leakage~\citep{kerriganDP,wu-etal-2024-mitigating-privacy}.
Data processing defenses, such as \emph{scrubbing}, remove PII from data to reduce leakage~\cite{pilan_text_2022,mosallanezhad_deep_2019,lukas2023analyzing}. 
Train time defenses such as differential privacy (DP) modify the models learning process to limit the influence of individual training examples while providing formal guarantees about the leakage~\citep{kerriganDP,li2022large,ponomareva-etal-2022-training,yu2022differentially}. Finally, post-training defenses, including model editing which remove or suppress PII-related knowledge from the LMs~\cite{wu-etal-2023-depn, chen-etal-2024-learnable,wu-etal-2024-mitigating-privacy}.
However, these defenses demonstrate poor privacy-utility trade-offs~\cite{lukas2023analyzing,wu-etal-2024-mitigating-privacy}, unintentionally increase susceptibility to other PII types~\cite{wu-etal-2024-mitigating-privacy,borkar2025privacy}, or can be circumvented by adversaries~\cite{xin2024a}.

To address these limitations, we propose \underline{P}rivacy-\underline{A}ware \underline{T}argeted \underline{C}ircuit Patc\underline{H}ing (\method), which allows editing relevant PII circuits to minimize leakage. 
We first apply mechanistic interpretability to discover the internal computational structures (or ``circuits'') in a LM responsible for PII leakage. We then ``patch'' or edit these circuits to mitigate the leakage.
Our contributions are:
\begin{enumerate}
\item A novel targeted circuit editing method, \method, that provides greater privacy-utility trade-off than existing defense mechanisms (\S\ref{sec:patch}, \S\ref{sec:results})
\item We identify and characterize the specific circuits responsible for leaking different PII types, like names, locations and race, revealing that distinct attention heads within these circuits that influence PII leakage.
(\S\ref{sec:piicircuit}).
\item An extensive ablation study of \method, further demonstrating its robustness across settings. (\S\ref{sec:discussions}).
\end{enumerate}
\section{Related Work}\label{sec:background}

\paragraph{PII Leakage in LMs.}
Memorized information from the training data can be shared via model weights from which an adversary can extract PII~\citep{lukas2023analyzing, yu_bag_2023, staab2024beyond, hayes_measuring_2025, hayes_measuring_2025-1}.
For instance, \citet{lukas2023analyzing} demonstrate PII leakage attacks against GPT2~\citep{radford2019language} models, where given access to LMs, adversaries can sample responses to extract sensitive information.
PII leakage vulnerabilities have been observed across diverse contexts, including pretrained models~\cite{nakka-etal,staab2024beyond,kim2023propile,panda2024teach,huang-etal-2022-large} and specialized applications \citep{hughes2024private, mireshghallah_trust_2024, xiao_large_2024}, indicating that privacy risks persist throughout the model's lifecycle and across diverse task domains.

\paragraph{Defenses against PII Leakage.}
A simple defense method is to remove PII from the data before training (``scrubbing'')~\cite{pilan_text_2022,mosallanezhad_deep_2019,lukas2023analyzing}.
However, such approaches are significantly expensive~\cite{wu-etal-2024-mitigating-privacy}, provide poor privacy-utility trade-offs~\cite{wu-etal-2023-depn,lukas2023analyzing,wu-etal-2024-mitigating-privacy}, and may allow an adversary to deduce personal attributes through auxiliary information~\cite{xin2024a}.
Alternatively, DP methods~\citep{feyisetan_privacy-_2020, kerriganDP, li2022large, shi_selective_2022, lee_private_2023} offer formal guarantees of privacy by injecting noise during training, effectively masking individual samples observed by the LM. Yet, this often results in a reduction in utility~\cite{lukas2023analyzing,wu-etal-2024-mitigating-privacy}.
Recent empirical defenses include identifying neurons that are responsible for memorizing PII and patching with steering vectors can reduce that memorization~\cite{wu-etal-2023-depn, chen-etal-2024-learnable, wu-etal-2024-mitigating-privacy}. However, \citet{wu-etal-2024-mitigating-privacy} indicate that such methods suffer from poor privacy-utility trade-offs due to limited components that can be edited, with the side effect of increasing the leakage of other PII types.

\section{\method: \underline{P}rivacy-\underline{A}ware \underline{T}argeted \underline{C}ircuit Patc\underline{H}ing}
\label{sec:patch}

\subsection{Problem Formulation}
Given a pre-trained model $\mathcal{M}$ and a private dataset $\mathcal{D}$ that contains PII $\mathcal{P}$, fine-tuning $\mathcal{M}$ on $\mathcal{D}$ results to a model $\hat{M}$ that is exposed to PII.  We also assume an adversary $\mathcal{A}$ with black-box access to $\hat{M}$. $\mathcal{A}$ seeks to infer specific PII types that are observable from $\mathcal{D}$ via prompting. Finally, a defense mechanism $\mathcal{DF}$ aims to reduce the effectiveness of $\mathcal{A}$ on extracting PII from $\hat{M}$, while maintaining the utility (i.e., performance) of $\mathcal{M}$~\citep{lukas2023analyzing}. $\mathcal{DF}$ can be applied before on $\mathcal{D}$, during fine-tuning on $\mathcal{D}$ or post-hoc, after training.

\subsection{Motivation}
We hypothesize that there is a set of unique elements (e.g., attention heads or outputs from heads to other parts of the Transformer block) within $\hat{M}$ responsible for PII leaks. By identifying and modifying these elements, we can reduce PII leakage while maintaining utility. 
More specifically, given a circuit discovery mechanism $\mathcal{CD}$, a private dataset $\mathcal{D}$, and a model $\mathcal{M}$, \method consists of the following three steps (detailed in Algorithm~\ref{alg:patch-overview}).

\begin{algorithm}[!t]
\small
\renewcommand{\algorithmiccomment}[1]{$\triangleright$ #1}
\begin{algorithmic}[1]
    \REQUIRE Model $\hat{M}$, PII types $P$, percentile threshold $p$, model editing method $A$, circuit discovery $CD$, private data $D$
    \ENSURE Privacy-enhanced model $\hat{M}$
    \FOR{each PII type $P_i \in P$}
        \STATE $S \gets \text{prompt\_builder}(D)$ \COMMENT{Generate prompts}
        \STATE $C_i \gets CD(\hat{M}, P_i, D, S)$ \COMMENT{Extract circuit}
        \STATE $\tau_i \gets ~\text{percentile}(\{s_e^{(i)} : e \in C_i\}, p)$ \\
        ~~~~~~\COMMENT{Compute Threshold}
        \STATE $E_i^{high} \gets ~\{e \in C_i : s_e^{(i)} \geq \tau_i\}$ \COMMENT{Select edges}
    \ENDFOR
    \STATE $E_{shared} \gets ~\bigcap_{i} E_i^{high}$ \COMMENT{Identify shared edges}
    \STATE $\hat{M} \gets A(\hat{M}, E_{shared})$ \COMMENT{Patch}
    \RETURN Modified model $\hat{M}$
\end{algorithmic}
\caption{PATCH: Privacy-Aware Targeted Circuit PatcHing}
\label{alg:patch-overview}
\end{algorithm}

\subsection{Step 1: Generate Prompts for Circuit Discovery.} 
To identify circuits responsible for PII leakage across all PII types, we employ Edge Attribution Patching with Integrated Gradients \citep[EAP-IG]{hanna2024have} as our circuit discovery mechanism $\mathcal{CD}$. EAP-IG is a mechanistic interpretability technique designed to efficiently discover circuits within Transformer models. We select EAP-IG for its effectiveness at identifying faithful circuits \citep{mueller2025mib}.

EAP-IG requires constructing prompts representing PII leakage so we can extract a circuit that is representative of that behavior. We construct pairs of ``clean'' and ``corrupt'' prompts. The clean version contains correct PII values and the corrupted version has these values replaced with alternatives from the same type. Using a private dataset $D$ that is tagged with a target PII type $P_i$, we select $1,000$ unique text spans containing that PII element $P_i$. The PII span is then replaced (``corrupted'') with PII that is semantically similar to that entity. Examples generated from a legal dataset \citep{chalkidis-etal-2019-neural} are shown in \autoref{tab:circuit-prompt-samples}.

\subsection{Step 2: Extract PII Leakage Circuits}
EAP-IG operates by comparing model behavior on clean prompts against corrupted variants where specific PII tokens have been altered. The method analyzes individual attention heads to measure two properties: (1) how sensitive each component's activations are to PII token corruption, and (2) how important each component is for correctly predicting PII values. Heads scoring high on both dimensions—being necessary for accurate PII prediction—are identified as part of the PII leakage circuit. Given $\hat{M}$ and $\mathcal{D}$, EAP-IG outputs a circuit $C_i$ containing \textit{nodes} that represent individual attention heads and \textit{edges} $\mathcal{e}$ that indicate information flow between nodes. Each node (attention head) and edge (connection to other nodes) receives an importance score $s_e^{(i)}$ quantifying their contribution to PII leaking behavior. In our experiments, we use the default hyperparameters recommended by \citet{hanna2024have}.

\begin{table}[!t]
    \centering
    \scriptsize
    \renewcommand{\arraystretch}{1.2}
    \begin{tabular}{p{1cm}|p{2.5cm}|p{2.5cm}}
    \toprule
        \textbf{PII Type} & \textbf{Original} & \textbf{Corrupted} \\
    \midrule
        Name & ``Mr. \colorbox{blue!10}{John} Smith vs. The State'' & ``Mr. \colorbox{red!10}{Heidi} Smith v. The State'' \\
        Location & ``The appellant was arrested in \colorbox{blue!10}{Berlin}.'' & ``The appellant was arrested in \colorbox{red!10}{New York}.'' \\
        Race & ``This case concerns a \colorbox{blue!10}{Romanian} national." & "This case concerns a \colorbox{red!10}{Turkish} national.'' \\
    \bottomrule
    \end{tabular}
    \caption{Examples from the European Court of Human Rights (ECHR) dataset \citep{chalkidis-etal-2019-neural}, illustrating PII types used in our original and corrupted prompts. \colorbox{blue!10}{Purple} represents a original token,  \colorbox{red!10}{red} represents a corrupted token.}
    \label{tab:circuit-prompt-samples}
\end{table}

\subsection{Steps 3 and 4: Compute Threshold and Perform Edge Selection}
We hypothesize that high-scoring edges indicate critical PII leakage pathways. Therefore, we aim to isolate those critical edges for patching.

Given a circuit discovery mechanism $\mathcal{CD}$ that has scored the edges from a circuit $C_i$ within $\hat{M}$, we require a mechanism to select the most influential edges. We first compute a threshold $\tau$ for each PII circuit $C_i$, given a specified percentile $p$. Next, we select the high-scoring edges whose scores are equal to or greater than the threshold $\tau$. We repeat this process for each PII type to obtain $\{E_i^{high}\}$ for all circuits $\{C_i\}$. Following prior work \citep{wang2025towards}, we select the high-scoring edges using percentile thresholds of $95\%$ and $99\%$.

\subsection{Step 5: Identify Shared PII Edges}
We aggregate edge scores across multiple PII types rather than treating each type independently,  assuming that PII leakage has common pathways throughout models. Therefore, we compute $E_{shared}$ as the intersection of high-scoring edges $E_i^{high}$ across all PII circuit types $C_i$.

\subsection{Step 6: Patch PII Edges}
Given a set of shared edges $E_{shared}$ from $\hat{M}$, we require a model editing mechanism $\mathcal{A}$ that alters those high-importance edges such that their influence on the model's behavior is reduced. We experiment with zero ablation, i.e.,setting edge weights to zero \citep{olah2020zoomin, pochinkov2024investigating}; and mean ablation, i.e.,replacing edge weights with their mean values \citep{chan2022causal, wang2023interpretability}, following their successful application in prior work \citep{bi2025unveiling}.

\section{Experimental Setup}
\label{sec:setup}

\subsection{PII Types and Private Data}
\label{sec:data}
We select \textit{names}, \textit{locations}, and \textit{race} as a set of PII types following prior work \citep{pilan_text_2022, hughes2024private, kim_generalizing_2024}.  
We use the European Court of Human Rights (ECHR) dataset \citep{chalkidis-etal-2019-neural}, which contains PII related to appellants and others involved in legal cases.  
To automatically identify PII across all train, validation and test sets, we apply the FLAIR name-entity-recognition (NER) tool.\footnote{FLAIR: \url{https://github.com/flairNLP/flair}} FLAIR achieves an accuracy of approximately $86\%$ \citep{yermilov_privacy-_2023}. Details of the NER label classes are provided in \autoref{tab:flair-classes}. Detailed corpus and PII statistics can be found in \autoref{app:data-analysis}.

\subsection{Circuit Discovery Prompts}

Using our private dataset, ECHR, that is tagged with the three PII types, we select unique text spans that containing the tagged PII element. We following prior work \citep{wang2025towards} and select 1,000 spans for each of the target PII types. The PII element in each span is then ``corrupted'' with PII that is semantically similar to that entity. We select entities using the faker library.\footnote{\url{https://faker.readthedocs.io/en/master/}} The clean and corrupted prompts are used to score circuit edges across all PII types (\autoref{tab:circuit-prompt-samples}).

\subsection{Base Models}

We use several open-weight LMs such as \GS~(117M), \GM~(345M), and \GL (774M) following prior work \citep{lukas2023analyzing, wu-etal-2024-mitigating-privacy}, as well as more recent models such as \LM \citep{grattafiori2024llama3herdmodels}, \QS and \QB \citep{qwen3technicalreport}.

\subsection{Fine-tuning Target Models}

To obtain LMs exposed to PII (target), we fine-tune all base models on the private ECHR dataset. We conduct our experiments using Hugging Face\footnote{\url{https://www.huggingface.co}} for all models. The max sequence length is set to $512$. All experiments on open-weight models are performed on one to four NVIDIA H100 GPUs. Fine-tuning uses a batch size of 8, the AdamW optimizer~\citep{loshchilov_decoupled_2019}, and a linear learning rate scheduler. Each baseline, DP and scrubbed defended model is trained for $4$ epochs.

\subsection{Adversary}

To emulate an adversary, we sample approximately four million tokens from each target LM with and without defenses. Each query begins with an empty prompt, then we issue 10,000 queries to each model, generating sequences of $256$ tokens from empty prompts using top-$k$ sampling with $k=40$ of which we apply random sampling.

To control for baseline PII leakage present in the base pretrained model, we establish a reference distribution by sampling 13 million tokens ($50,000$ queries) from it prior to fine-tuning. Any PII instances appearing in this baseline are excluded from our leakage measurements, ensuring we only attribute leakage to the fine-tuning process rather PII aquired during pretraining. We repeat the attack three times to quantify variance. This is following prior work in PII leakage \citep{lukas2023analyzing}.

\subsection{Metrics}
\paragraph{Privacy Leakage.} To assess the adversary success in leaking PII, we directly compare verbatim LM outputs against FLAIR annotations, which serve as ground truth.
\emph{Precision} measures the proportion of PII in the model's output is PII that was also present in the training data. \emph{Recall} indicates how much of the total PII observable in the training data is exposed.

\paragraph{Faithfulness.} We assess how accurately a discovered circuit represents the causal mechanism underlying PII leakage. 
A circuit is considered faithful if ablating any components outside the circuit does not affect the model's performance, indicating that the circuit alone explains the behavior~\citep{prakash2023fine, hanna2024have}. We report the normalized faithfulness as: 

$$\frac{P_{\text{method}} - P_{\text{corrupted}}}{P_{\text{baseline}} - P_{\text{corrupted}}}$$

\noindent $P_{\text{method}}$ is the performance of a circuit after applying EAP-IG, $P_{\text{baseline}}$ is the original model performance, and $P_{\text{corrupted}}$ is the corrupted baseline performance. 
The score ranges from $0$ (no resemblance to the original model) to $1$ (full recovery of performance).

\paragraph{Circuit Overlap.} To identify how PII circuits interact and where editing those circuits is optimal, we require a circuit overlap measure. Following \citet{hanna2024have}, we use the \emph{overlap metric} based on high scoring circuit elements that meet a percentile threshold. For each PII type $P_i \in T$, we extract circuit $C_i$ with node or edge scores $s_e^{(i)}$ using EAP-IG $\mathcal{CD}$. We then compute a threshold $\tau_i$ and identify high-importance edges $E_i^{high}$. The overlap between two PII circuits $C_i$ and $C_j$ is calculated using the Jaccard index:
$$\text{Overlap}(C_i, C_j; \tau) = \frac{|E_i^{high} \cap E_j^{high}|}{|E_i^{high} \cup E_j^{high}|} \times 100\%$$

\noindent $\tau$ denotes the percentile threshold. High overlap indicates shared computational pathways across PII types. We analyze edge overlap at percentile thresholds of $p \in \{95, 99\}$ following \citep{wang2025towards}, with higher thresholds selecting fewer but more critical edges for editing.

\begin{figure*}[!t]
    \centering
    \includegraphics[width=0.92\linewidth]{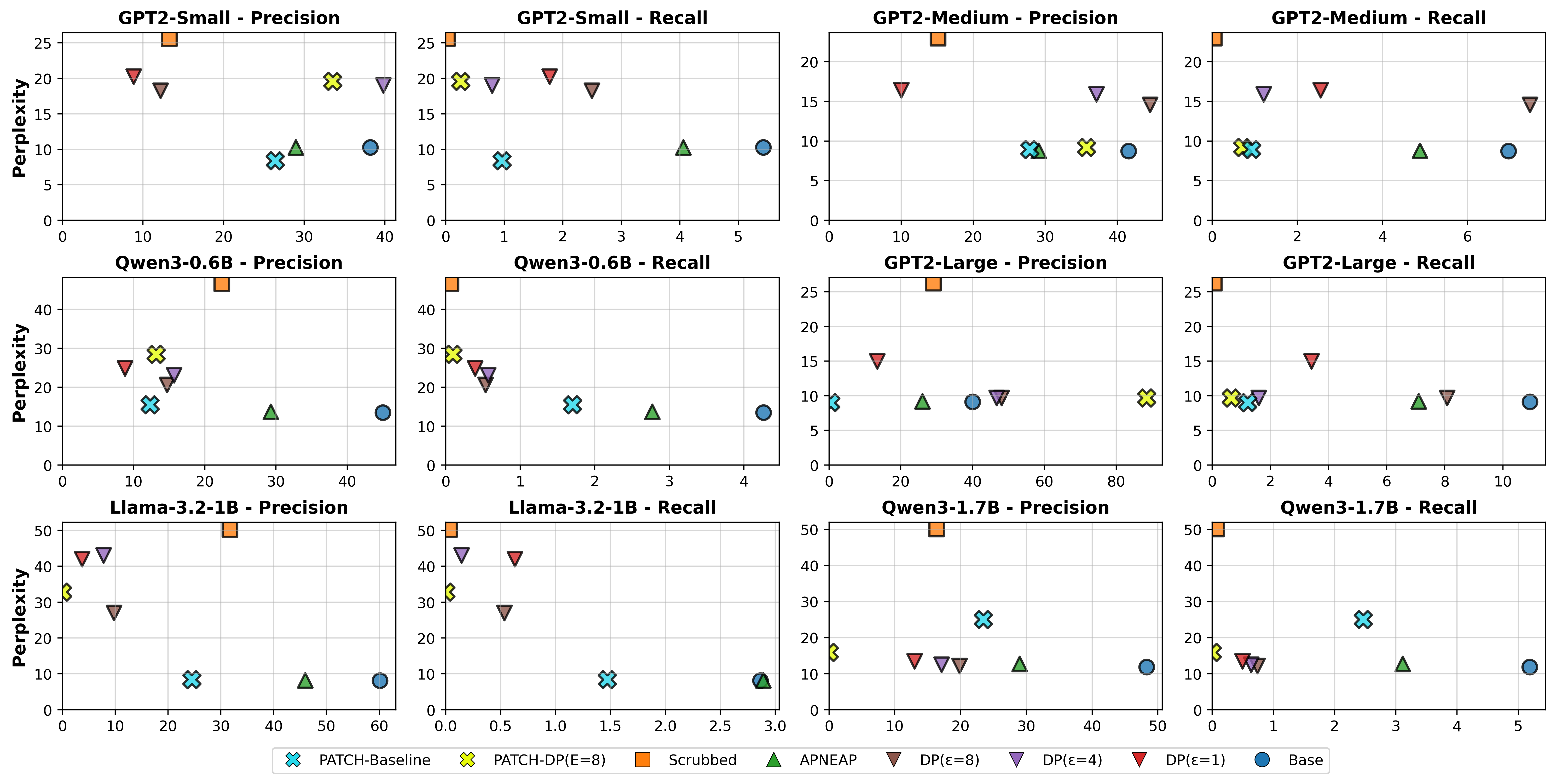}
    \caption{\textbf{Comparison of \method with other defenses:}  lower perplexity, precision and recall scores are preferred (closer to lower left corner).}
    \label{fig:final-methods-comparision}
\end{figure*}

\paragraph{Utility.} We use \emph{perplexity} over the ECHR test set, similar to prior work~\citep{lukas2023analyzing, wu-etal-2023-depn, chen-etal-2024-learnable, wu-etal-2024-mitigating-privacy}, to evaluate the impact on LM utility before and after applying \method and baseline defense mechanisms.

\subsection{Defense Baselines}

We compare with the following defenses:

\paragraph{APNEAP.} Augmented Privacy Neuron Editing via Activation Patching~\cite{wu-etal-2024-mitigating-privacy} identifies and ablates individual neurons responsible for PII leakage. This is the current state-of-the-art editing-based defense which outperforms prior editing approaches~\cite{wu-etal-2023-depn,chen-etal-2024-learnable}.

\paragraph{DP.} We fine-tune LMs using differentially private stochastic gradient descent \citep[DP-SGD]{10.1145/2976749.2978318}. We consider different privacy loss parameter $\epsilon$ =\{8, 4, 1\} where a lower $\epsilon$ indicates stronger privacy guarantee. For DP training, we use the fastDP library \citep{bu2022differentially, bu2023differentially}. Following prior work \citep{lukas2023analyzing}, each model is trained using DP-SGD for 4 epochs using ($\epsilon$, $\delta$) = ($\{8,4,1\}$, $1N)$ where $N$ is the size of the training dataset. We use a maximum per-sample gradient norm of $1$.

\paragraph{Scrubbing.} We first remove PII from the data, and then use it for fine-tuning LMs. All information related to the selected PII types (names, locations, race) is removed. Again, we use an NER tool, FLAIR, to identify any spans containing PII, we then redact those spans. Redaction means replacing the identified span with a masking token. Models are then trained on the resulting scrubbed documents.

\paragraph{\method.} We evaluate two variants of our proposed \method approach. First, \method-Baseline applies our method to a model $\mathcal{M}$ trained without any existing privacy defenses. Second, \method-DP($\epsilon=8$) applies our method to a model $\mathcal{M}$ that has been fine-tuned with DP($\epsilon=8$).

\section{Results}
\label{sec:results}

\autoref{fig:final-methods-comparision} presents privacy-utility tradeoffs for \method against three baseline defenses across all models. We measure PII leakage with precision and recall, and utility using perplexity. We evaluate two variants of our method: \textsc{\method-Baseline} applied to non-private models, and \textsc{\method-DP}($\epsilon$) applied to a model fine-tuned with DP at $\epsilon=8$.\footnote{We ran experiments with a DP fine-tuned model at $\epsilon=\{4, 1\}$, however we obtained poor utility. Table of experiments is available in \autoref{app:all-baseline-results}.}

\paragraph{\method-Baseline achieves strong privacy-utility tradeoffs.}
Across all models, \textsc{\method-Baseline} reduces PII extraction precision by $40\%$--$90\%$ relative to undefended baselines. For example, \LM precision decreases from $60\%$ to $0.5$, while incurring only modest utility costs; \GM perplexity increases from approximately $9$ to $10$. Recall consistently decreases by $80\%$--$86\%$ across architectures, with \GM decreasing from $8.5\%$ to $1.2\%$ and \GL from approximately $11\%$ to $1.8\%$ while maintaining a utility within $1$ point of the baseline perplexity. These reductions demonstrate that PII leakage localizes to identifiable circuit edges rather than distributing across the entire parameter space. The consistency across architectures indicates that PII leaks via specific mechanisms.

\paragraph{\method-DP provides maximal privacy at varied utility costs.}
Combining circuit patching with DP yields the strongest privacy protection, achieving recall $<1\%$ across all models, but with model-dependent utility impact. For smaller models, \method-DP($\epsilon=8$) maintains reasonable utility: on \GS, perplexity increases to $19.5$ (vs. $18.3$ for DP($\epsilon=8$) alone, and $10.3$ for Base), while recall drops to $<0.5\%$ compared to $2.0\%$ for DP alone. Similar patterns emerge for \LM where perplexity is $32$ compared to $27$ for the DP baseline, however recall approaches $0\%$. This demonstrates good protection, but fails to maintain utility.

\paragraph{Be careful of precision increases.} An important finding in \method-DP($\epsilon=8$) in \GL, is that it successfully reduces recall, however it exhibits concerning spikes in precision with stable perplexity, particularly evident in \GL where precision reaches $\sim80$ while recall drops to $\sim1$. This tells us the leaks have become highly accurate. For adversaries mounting training data extraction attacks, high precision and low recall suggest a vulnerability: \textit{the model may leak infrequently, but successful extractions yield genuine, trustworthy PII.} This pattern is typically present in the baseline model. \method-Baseline corrects this issue, however \textit{it is important to note that a defense may simultaneously increase model susceptibility}.

\paragraph{Comparison to existing defenses.}
Undefended baseline models exhibit substantial PII leakage, with precision ranging from $38.6\%$ to $60.8\%$ and recall from $3.0\%$ to $11.0\%$ across architectures, confirming the severity of leaks in models fine-tuned on sensitive data. Among existing defense mechanisms, DP provides strong privacy protections, reducing recall to $\leq6\%$ across all models. For example, \LM decreases to $0.5\%$. However, these models incur substantial utility costs with perplexity increases of $1.5$ to $7.9$ times the baseline perplexity. APNEAP~\citep{wu-etal-2024-mitigating-privacy} maintains utility within $+1.0$ perplexity of baseline models but provides limited privacy protection, reducing extraction precision by only $\sim10$-$18$\% and recall by $\sim1$-$3$\%.

\paragraph{Practical implications.}
Overall, our results demonstrate that circuit-based interventions can provide quantifiable privacy-utility tradeoffs complementary to formal DP guarantees. For deployment scenarios where moderate privacy protection suffices and utility is paramount, \method-Baseline offers $\sim1.5$-$8$\% reduction in recall with minimal perplexity overhead. For high-security applications requiring maximal privacy, \method-DP($\epsilon=8$) can achieve $\le1\%$ in precision and recall, though \textit{practitioners should be conscious of utility}, as there were cases, \QB, where utility was not preserved. \textit{These differences across models underscores the need for careful evaluation of defense mechanism impacts across model architectures}.

\paragraph{Privacy budget considerations.} Our circuit identification process analyzes model activations on the training data to locate relevant circuits for PII leakage. This data-dependent analysis is not accounted for in the privacy hyperparameter ($\epsilon$) that was specified during DP fine-tuning. For scenarios requiring strict formal guarantees, circuit identification should be performed on a separate dataset, after which patching operations would preserve the training-time privacy guarantees. \textit{In this work, we demonstrate valuable empirical privacy protection through targeted circuit interventions.}

\section{Analysis of PII Leakage Circuits}
\label{sec:piicircuit}

\begin{figure*}[!t]
    \centering
    \includegraphics[width=0.95\linewidth]{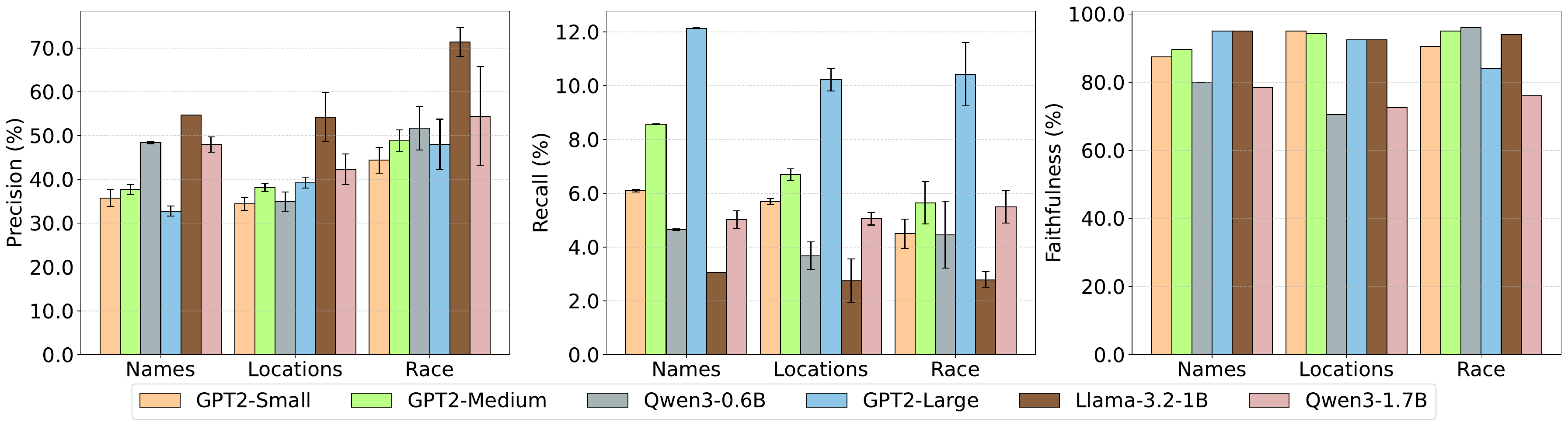}
    \caption{\textbf{Results for PII leakage and faithfulness:} we present the precision (\textbf{left}), and recall (\textbf{middle}) of the PII extracted, and finally, the faithfulness of the discovered PII circuits (\textbf{right}).}
    \label{fig:baselines-summary}
\end{figure*}

\subsection{PII leakage and Circuits} 
\label{subsec:faithful-circuits}
\autoref{fig:baselines-summary} presents the results of the PII leakage and circuit discovery, for three PII types: names, location, and race. 

We find precision is variable with regard to model architectures, however, within architectures, such as GPT2 and Qwen3, recall increases with model size. This corroborates prior work \citep{lukas2023analyzing}. 
Among other models, \LM~has the highest precision indicating that the PII produced by this model is more likely to be from the training data. In contrast, \GL~has the highest recall exposing the most PII in training data. 
Furthermore, the PII leakage, as seen with precision and recall, varies across different PII types and across models.
Regardless of the model or PII type faithfulness is high, indicating that we can reliably identify the circuits for PII leakage. 
Faithfulness scores (above $75\%$) are similar to prior circuit discovery work~\citep{wang2025towards}.
\textit{Overall, we can reliably identify circuits for PII leakage, and this remains high across all models.}\footnote{We also analyzed circuit overlaps across defense baselines. Overlap metrics are available in \autoref{app:model-overlap-analysis}.}

\subsection{PII Circuit Overlaps}
\label{subec:overlaps}

\autoref{tab:circuit-overlap} shows the results for PII circuit overlap. This allows us to understand the structure of the individual PII circuits, and how editing these may impact an attack.

\begin{table}[!h]
\centering
\scriptsize
\setlength{\tabcolsep}{4pt}
    \begin{tabular}{lccc}
        \toprule
            \multirow{2}{*}{\textbf{Model}} & \multicolumn{3}{c}{\textbf{PII Circuit Overlaps (nodes/edges ($\%$))}} \\
        \cmidrule(lr){2-4}
        & \textbf{Name-Location} & \textbf{Name-Race} & \textbf{Location-Race} \\
        \midrule
            \GS & 79 / 39 & 88 / 48 & 77 / 38 \\
            \GM & 66 / 23 & 71 / 26 & 72 / 24 \\
            \QS & 62 / 50 & 61 / 41 & 54 / 41 \\
            \GL & 70 / 30 & 69 / 24 & 65 / 24 \\
            \LM & 82 / 44 & 82 / 40 & 84 / 40 \\
        \QB & 87 / 25 & 87 / 24 & 96 / 49 \\
        \bottomrule
    \end{tabular}
    \caption{\textbf{Results of circuit overlap analysis across PII types and models:} we present \emph{circuit overlap ($\%$)} between faithful PII circuits (format: ``nodes/edges'').}
    \label{tab:circuit-overlap}
\end{table}

\paragraph{Minimal edge overlap across PIIs types.} Comparing connectivity patterns of circuits responsible for different PII types reveals substantial differences in the edge overlap. We observe a consistently low overlap between the edges of PII circuits across all models. \GS shows $38.8\%$ overlap between race and names, $37.6\%$ between race and locations, and $48.3\%$ between names and locations, while \GL and \QB exhibit even lower overlap of $\sim25\%$. 
This suggests that different PII circuits have distinct pathways. Low edge overlap combined with high node overlap in larger models indicates greater circuit specialization with increased capacity. This also explains the observations in prior node ablation work~\cite{wu-etal-2023-depn,wu-etal-2024-mitigating-privacy,borkar2025privacy} where minimizing leakage of one PII through node editing, increases others. \textit{This pattern also suggests that LMs do not maintain entirely separate mechanisms for retrieving PII, but instead rely on a smaller set of shared attention patterns across PII types.}

\begin{figure*}[!t]
    \centering
    \includegraphics[width=\linewidth]{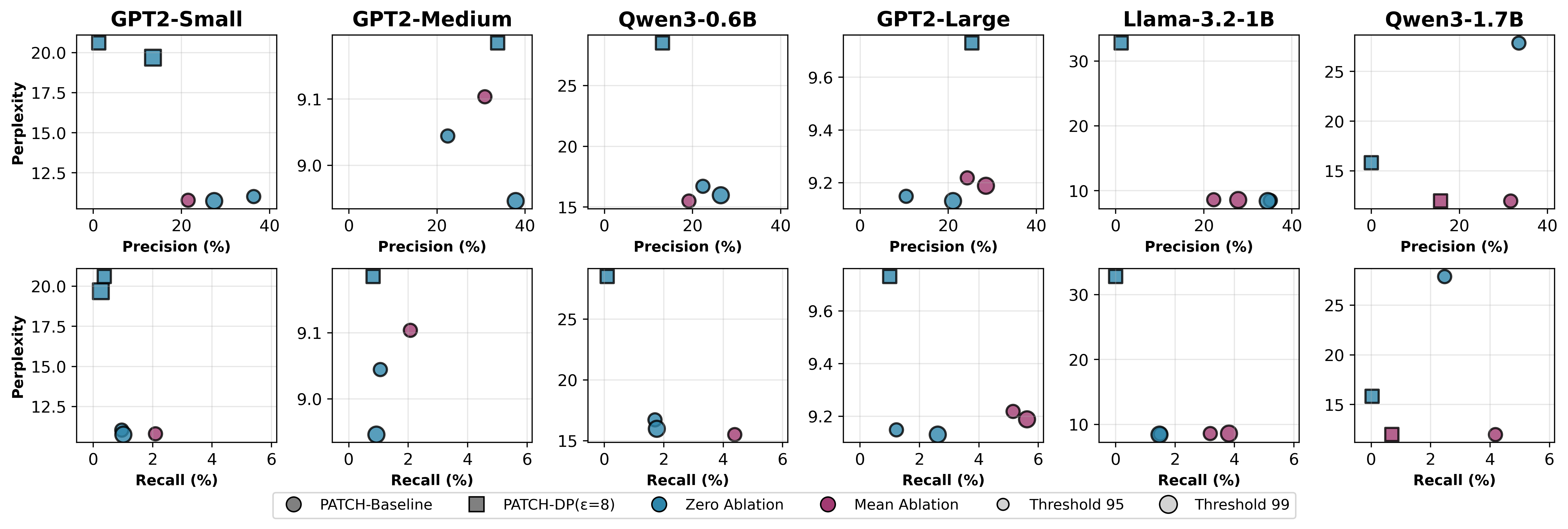}
    \caption{\textbf{Results of \method across varying hyperparameters:} we compare \method-Baseline with \method-DP($\epsilon=8$) with alternating ablation strategies, zero and mean, and edge thresholds, $95$ and $99$.}
    \label{fig:hyperp-recall-analysis}
\end{figure*}

\begin{figure}[!t]
    \centering
    \includegraphics[width=\linewidth]{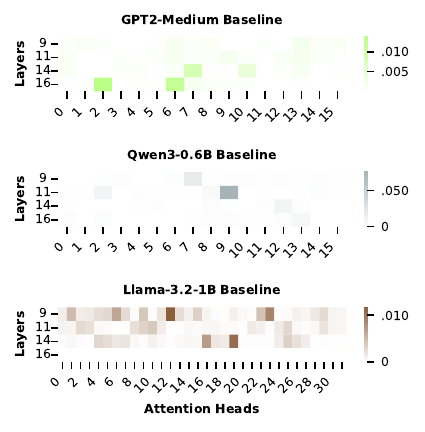}
    \caption{\textbf{Influential PII Circuit Components:} average EAP-IG scores for each attention head in each layer, across all identified PII circuits.}
    \label{fig:pii-components-analysis-subset}
\end{figure}

\subsection{Influential Edges in PII Circuits}
\label{subec:influence}

Given the low overlapping scores observed among edges, we investigate their individual influence. We observe the scores generated by the circuit discovery method (EAP-IG). We present heatmaps for a subset of the models attention layers and heads in \autoref{fig:pii-components-analysis-subset}.\footnote{See full heatmaps of each model's attention layers and heads in \autoref{app:pii-components-analysis-all}.}
We find that circuit discovery finds distinct influence patterns of attention head edges across architectures. \GM displays a more distributed pattern with notable activation in layers $14$-$16$ across multiple heads. \LM shows sparse but intense activation primarily in later layers, with early layer concentration with high scores in \emph{layer $9$}. 
This is also visible in \QS and \QB where \emph{layer $0$} is highly influential. Interestingly, some layer and heads are very pronounced. We identify \emph{layer $11$ head $9$} as highly influential in both \textit{Qwen3} models.
\textit{Overall, key edges vary across architectures but systematic circuits that drive PII leakage exist across all models.}

\section{Ablation Study}
\label{sec:discussions}

To be better understand \method, we perform attacks under different hyperparameter settings. We report the results in \autoref{fig:hyperp-recall-analysis}. As indicated in \autoref{sec:patch}, we use $95\%$ and $99\%$ percentile threshold on the importance scores of the edges. We also evaluate \method with \textit{zero} and \textit{mean} ablation with both thresholds. This helps evaluate whether there are other configurations which can improve privacy-utility trade-offs.

\paragraph{Model editing strategies show no universal differences.} Our comparative analysis of mean versus zero ablation strategies reveals no systematic preference across model architectures and scales. For smaller models such as \GS, zero ablation configurations cluster in regions of lower perplexity with marginally reduced F1-scores, suggesting minimal functional differentiation between ablation methods. \GM and \QB mean and zero ablation variants are without clear separation. Practitioners should evaluate both strategies during post-training. Optimal ablation appears model specific.

\paragraph{Threshold selection has model-specific effects.} The thresholds $95\%$ and $99\%$ comparison reveals model-specific patterns without consistent trends. \GS shows threshold 99 configurations incurring higher perplexity costs for equivalent privacy gains, while \GL exhibits the opposite relationship where stricter thresholds maintain comparable utility. \QS and \QB architectures demonstrate negligible threshold sensitivity, with both settings producing similar privacy-utility trade-offs. This heterogeneity indicates that optimal threshold selection requires model-specific calibration.

\section{Conclusion}
We introduced \method, a method grounded in mechanistic interpretability for mitigating PII leakage in LMs through targeted editing. Our approach identifies shared computational pathways responsible for PII memorization across PII types and selectively ablates high-importance edges to reduce privacy risks. Empirical results demonstrate that \method achieves superior privacy-utility trade-offs compared to prior work. Our extensive experiments verify the effectiveness of \method across multiple model scales and PII categories compared to non-private fine-tuned baselines. These findings demonstrate that targeted circuit interventions can provide effective privacy protection while preserving the model's core computational pathways.

\section*{Limitations}
In this study, we empirically demonstrate that our method substantially improves open-weight LMs in mitigating PII leaks, however we acknowledge our evaluation is limited to a subset of PII types. We hope to extend our proposed method to a broader set of PII in the future. Finally, experiments have not been conducted on large models, due to circuit discovery requiring intensive resources and time, however as new methods arise they can integrate into our proposed method.

\section*{Acknowledgments}
AH is supported by the Centre for Doctoral Training in Speech and Language Technologies (SLT) and their
Applications funded by UK Research and Innovation [EP/S023062/1]. This work is supported in part by Intel (in the context of Private AI consortium), and the Government of Ontario. Vasisht is supported by David R. Cheriton Scholarship, Cybersecurity and Privacy Excellence Graduate Scholarship, and an IBM PhD Fellowship. Additionally, we acknowledge IT Services at The University of Sheffield for the provision of services for High Performance Computing. Finally, views expressed in the paper are those of the authors and do not necessarily reflect the position of the funding agencies.

\bibliography{paper}

\appendix
\newpage 
\section{Dataset Analysis}
\label{app:data-analysis}

\autoref{tab:echr-stats} presents detailed statistics regarding the corpora used in our experiments.

\begin{table}[h]
    \centering
    \renewcommand{\arraystretch}{1.3} 
    \resizebox{\columnwidth}{!}{%
    \begin{tabular}{lcccccc}
        \toprule
        \multicolumn{2}{c}{} & \multicolumn{2}{c}{\textbf{Words}} & \multicolumn{3}{c}{\textbf{PII}} \\
        \midrule
        \textbf{Task} & \textbf{Tr/Dev} & \textbf{Mean} & \textbf{Max} & \textbf{Names} & \textbf{Locations} & \textbf{Race} \\
        \midrule
        ECHR & 118161/26258 & 70 & 8723 & 63685 & 2089 & 9921 \\
        \bottomrule
    \end{tabular}%
    }
    \caption{
    Distribution of source documents in ECHR. The mean and maximum word count for all source documents is presented along with an overview of the quantity of PII in all documents.
    }
    \label{tab:echr-stats}
\end{table}


\section{Flair Classes}
In \autoref{tab:flair-classes}, we present the classes of PII used to tag and monitor for leaks in our experiments.

\begin{table}[!h]
    \centering
    \scriptsize
    \renewcommand{\arraystretch}{1.2}
    \begin{tabular}{p{1cm}|p{1.5cm}|p{3cm}}
    \toprule
         Class & Description & Example from ECHR \\
     \midrule
         PERSON     & Names of persons    & According to the Government, \colorbox{red!10}{Mr L.} had submitted that on the date in question. \\
         LOC        & General locations   & Filed a letter with the Chancellor of the \colorbox{red!10}{Jagiellonian University} in \colorbox{red!10}{Kraków}. \\
         NORP       & Race, national, religious groups & a group of \colorbox{red!10}{Turkish} nationalists \\
     \bottomrule
    \end{tabular}
    \caption{The classes used by the Flair tagger for word classification. We highlight the identified span in \colorbox{red!10}{red}}
    \label{tab:flair-classes}
\end{table}

\newpage
\section{Results of all baseline defenses}
\label{app:all-baseline-results}
In \autoref{tab:summary} we present the results of all defense baselines.

\definecolor{mygreen}{HTML}{C8E6C9}
\definecolor{myorange}{HTML}{FFECB3}
\definecolor{myred}{HTML}{FFCDD2}
\definecolor{gray!15}{HTML}{F0F0F0}

\setlength{\tabcolsep}{0.5pt}

\begin{table}[!h]
    \centering
    \small
    \begin{tabular}{l|ccc}
        \toprule
        \textbf{Defense} & \textbf{Perpl $\downarrow$} & \textbf{Prec (\%) $\downarrow$} & \textbf{Rec (\%) $\downarrow$} \\
        \midrule
        \multicolumn{4}{c}{\textbf{GPT-Small}} \\
        \midrule
        \cellcolor{gray!15}\textbf{Baseline} & \cellcolor{gray!15}10.26 & \cellcolor{gray!15}38.19 $\pm$ 4.98 & \cellcolor{gray!15}5.43 $\pm$ 0.80 \\
        \textbf{APNEAP} & \cellcolor{myred}10.27 & \cellcolor{mygreen}29.92 $\pm$ 3.70 & \cellcolor{myorange}5.02 $\pm$ 0.87 \\
        \textbf{DP ($\epsilon$=8)} & \cellcolor{myred}18.27 & \cellcolor{mygreen}10.38 $\pm$ 4.60 & \cellcolor{mygreen}2.27 $\pm$ 0.74 \\
        \textbf{DP ($\epsilon$=4)} & \cellcolor{myred}19.01 & \cellcolor{mygreen}11.49 $\pm$ 1.47 & \cellcolor{mygreen}2.06 $\pm$ 0.35 \\
        \textbf{DP ($\epsilon$=1)} & \cellcolor{myred}20.27 & \cellcolor{mygreen}8.01 $\pm$ 4.22 & \cellcolor{mygreen}2.00 $\pm$ 0.76 \\
        \midrule
        \multicolumn{4}{c}{\textbf{GPT2-Medium}} \\
        \midrule
        \cellcolor{gray!15}\textbf{Baseline} & \cellcolor{gray!15}8.72 & \cellcolor{gray!15}41.56 $\pm$ 5.55 & \cellcolor{gray!15}6.97 $\pm$ 1.46 \\
        \textbf{APNEAP} & \cellcolor{myred}8.76 & \cellcolor{mygreen}27.01 $\pm$ 3.34 & \cellcolor{mygreen}4.53 $\pm$ 0.79 \\
        \textbf{DP ($\epsilon$=8)} & \cellcolor{myred}14.91 & \cellcolor{mygreen}36.13 $\pm$ 8.95 & \cellcolor{myorange}6.12 $\pm$ 2.17 \\
        \textbf{DP ($\epsilon$=4)} & \cellcolor{myred}15.90 & \cellcolor{mygreen}32.47 $\pm$ 6.40 & \cellcolor{mygreen}1.04 $\pm$ 0.23 \\
        \textbf{DP ($\epsilon$=1)} & \cellcolor{myred}16.40 & \cellcolor{mygreen}9.51 $\pm$ 3.17 & \cellcolor{mygreen}3.04 $\pm$ 1.21 \\
        \midrule
        \multicolumn{4}{c}{\textbf{Qwen3-0.6B}} \\
        \midrule
        \cellcolor{gray!15}\textbf{Baseline} & \cellcolor{gray!15}13.48 & \cellcolor{gray!15}45.01 $\pm$ 8.39 & \cellcolor{gray!15}4.26 $\pm$ 0.49 \\
        \textbf{APNEAP} & \cellcolor{myred}14.54 & \cellcolor{mygreen}29.26 $\pm$ 1.61 & \cellcolor{mygreen}2.77 $\pm$ 0.38 \\
        \textbf{DP ($\epsilon$=8)} & \cellcolor{myred}20.63 & \cellcolor{mygreen}15.69 $\pm$ 0.00 & \cellcolor{mygreen}0.57 $\pm$ 0.00 \\
        \textbf{DP ($\epsilon$=4)} & \cellcolor{myred}22.10 & \cellcolor{mygreen}14.83 $\pm$ 2.72 & \cellcolor{mygreen}0.55 $\pm$ 0.10 \\
        \textbf{DP ($\epsilon$=1)} & \cellcolor{myred}24.84 & \cellcolor{mygreen}8.78 $\pm$ 0.67 & \cellcolor{mygreen}0.39 $\pm$ 0.08 \\
        \midrule
        \multicolumn{4}{c}{\textbf{GPT2-Large}} \\
        \midrule
        \cellcolor{gray!15}\textbf{Baseline} & \cellcolor{gray!15}9.14 & \cellcolor{gray!15}40.02 $\pm$ 7.62 & \cellcolor{gray!15}10.92 $\pm$ 0.96 \\
        \textbf{APNEAP} & \cellcolor{myred}9.63 & \cellcolor{mygreen}26.01 $\pm$ 4.06 & \cellcolor{mygreen}7.10 $\pm$ 0.56 \\
        \textbf{DP ($\epsilon$=8)} & \cellcolor{myred}12.00 & \cellcolor{myorange}42.05 $\pm$ 0.78 & \cellcolor{mygreen}5.38 $\pm$ 0.14 \\
        \textbf{DP ($\epsilon$=4)} & \cellcolor{myred}12.69 & \cellcolor{mygreen}37.06 $\pm$ 1.38 & \cellcolor{mygreen}2.76 $\pm$ 0.25 \\
        \textbf{DP ($\epsilon$=1)} & \cellcolor{myred}14.97 & \cellcolor{mygreen}15.32 $\pm$ 0.00 & \cellcolor{mygreen}1.97 $\pm$ 0.00 \\
        \midrule
        \multicolumn{4}{c}{\textbf{Llama-3.2-1B}} \\
        \midrule
        \cellcolor{gray!15}\textbf{Baseline} & \cellcolor{gray!15}8.13 & \cellcolor{gray!15}60.12 $\pm$ 8.59 & \cellcolor{gray!15}2.86 $\pm$ 0.16 \\
        \textbf{APNEAP} & \cellcolor{myred}8.20 & \cellcolor{mygreen}45.96 $\pm$ 6.80 & \cellcolor{myorange}2.89 $\pm$ 1.44 \\
        \textbf{DP ($\epsilon$=8)} & \cellcolor{myred}15.66 & \cellcolor{mygreen}9.76 $\pm$ 0.92 & \cellcolor{mygreen}0.53 $\pm$ 0.11 \\
        \textbf{DP ($\epsilon$=4)} & \cellcolor{myred}20.99 & \cellcolor{mygreen}5.75 $\pm$ 1.14 & \cellcolor{mygreen}0.35 $\pm$ 0.10 \\
        \textbf{DP ($\epsilon$=1)} & \cellcolor{myred}42.01 & \cellcolor{mygreen}1.16 $\pm$ 0.92 & \cellcolor{mygreen}0.10 $\pm$ 0.06 \\
        \midrule
        \multicolumn{4}{c}{\textbf{Qwen3-1.7B}} \\
        \midrule
        \cellcolor{gray!15}\textbf{Baseline} & \cellcolor{gray!15}11.89 & \cellcolor{gray!15}48.26 $\pm$ 6.06 & \cellcolor{gray!15}5.19 $\pm$ 0.23 \\
        \textbf{APNEAP} & \cellcolor{myred}12.76 & \cellcolor{mygreen}28.96 $\pm$ 3.32 & \cellcolor{mygreen}3.11 $\pm$ 0.23 \\
        \textbf{DP ($\epsilon$=8)} & \cellcolor{myred}12.15 & \cellcolor{mygreen}20.43 $\pm$ 15.03 & \cellcolor{mygreen}0.78 $\pm$ 0.36 \\
        \textbf{DP ($\epsilon$=4)} & \cellcolor{myred}12.50 & \cellcolor{mygreen}17.10 $\pm$ 15.03 & \cellcolor{mygreen}0.63 $\pm$ 0.45 \\
        \textbf{DP ($\epsilon$=1)} & \cellcolor{myred}13.45 & \cellcolor{mygreen}13.00 $\pm$ 12.46 & \cellcolor{mygreen}0.49 $\pm$ 0.43 \\
        \bottomrule
    \end{tabular}    
    \caption{\textbf{Impact of DP Fine-tuning}: we use perplexity (``Perpl'') for utility, precision (``Prec'') and recall (``Rec'') for PII leakage, and normalized faithfulness (``Faith''), averaged across all PII types. $\downarrow$ ($\uparrow$) indicates lower (higher values) is preferred. We use \colorbox{gray!15}{Gray} for the baseline (no defenses), \colorbox{mygreen}{Green} if better, \colorbox{myred}{Red} if worse, and \colorbox{myorange}{Orange} if within standard deviation of the baseline.}
    \label{tab:summary}    
\end{table}







\clearpage
\section{Influential Circuit Components}
\label{app:pii-components-analysis-all}

\begin{figure*}[!t]
    \centering
    \includegraphics[width=0.95\linewidth]{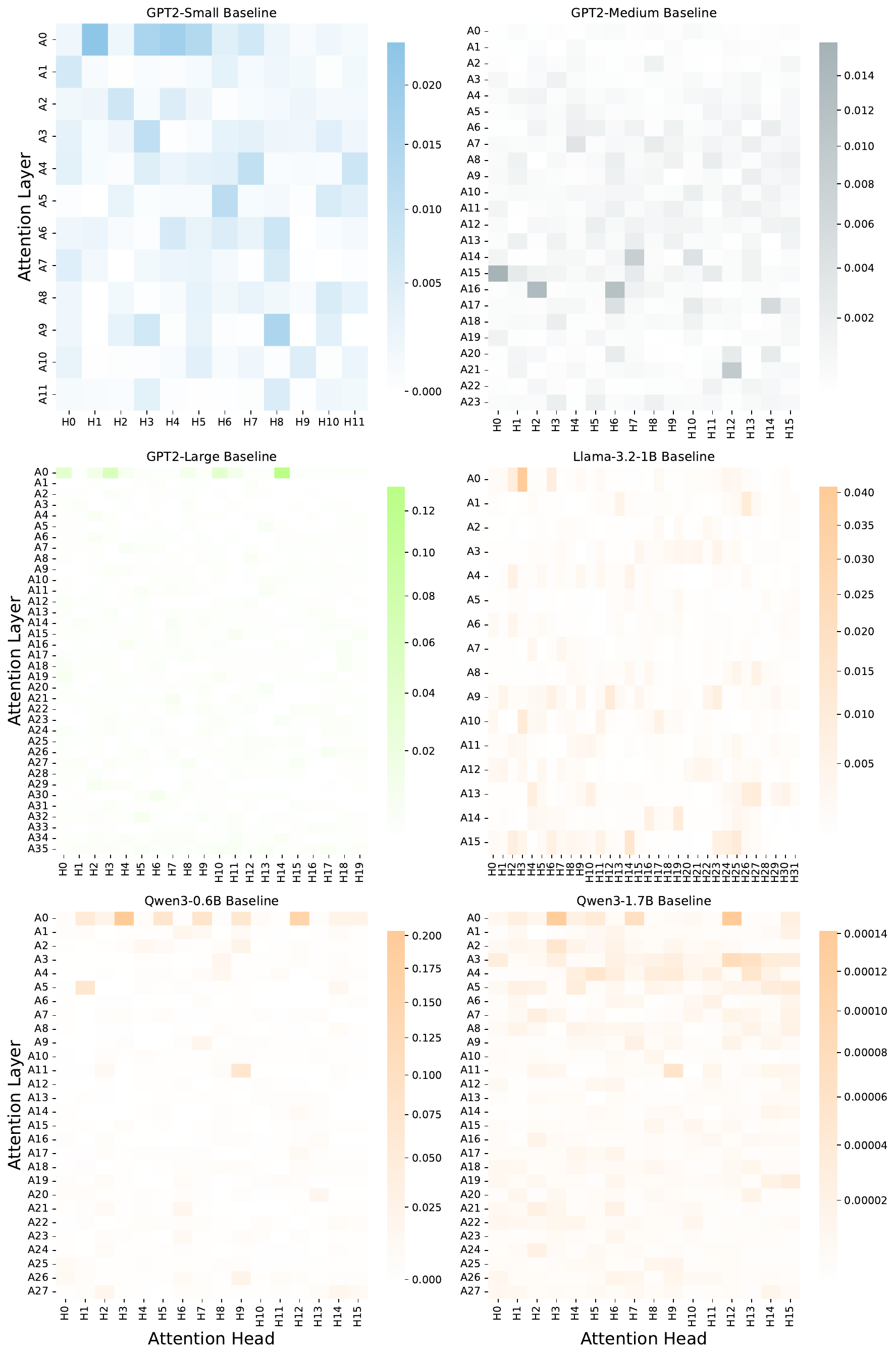}
    \caption{\textbf{Influential PII Circuit Components:} Heatmaps show the average EAP-IG scores for each attention head within each layer, across all identified PII circuits.}
    \label{fig:pii-components-analysis-all-app}
\end{figure*}

In order to understand more about PII leakage within the attention layers of models we displays EAP-IG scored attention layers where those layers score above a 95\% threshold across all scored components. We present our analysis in \autoref{fig:pii-components-analysis-all-app}.

\section{Circuit analysis of defenses.}
\label{app:model-overlap-analysis}

\begin{figure*}[!t]
    \centering
    \includegraphics[width=\linewidth]{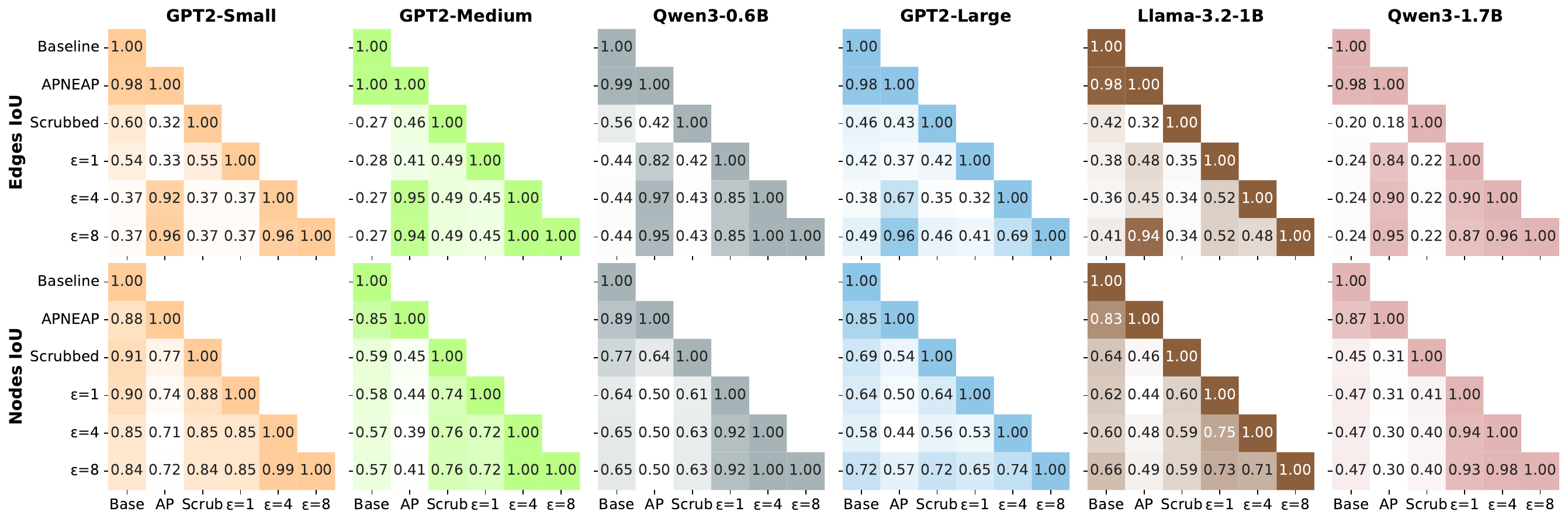}
    \caption{\textbf{Circuit overlap across defenses:} percentage overlap of edges (\textbf{upper}) and nodes (\textbf{lower}) in PII circuits.}
    \label{fig:node-edge-overlap-models-comparison}
\end{figure*}

In \autoref{fig:node-edge-overlap-models-comparison}, we present the circuit overlap between models in trained on differing defenses.

\end{document}